\documentclass[runningheads]{llncs}
\usepackage[T1]{fontenc}
\usepackage{graphicx}
\usepackage{cite}
\usepackage{amsmath,amssymb,amsfonts}
\usepackage{algorithmic}
\usepackage{graphicx}
\usepackage{textcomp}
\usepackage{xcolor}
\usepackage{comment}
\usepackage{pgfplots}
\pgfplotsset{compat=1.18}
\usepackage{booktabs}
\usepackage{multirow}
\usepackage{pifont}     
\usepackage{listings}

\newcommand{\red}[1]{\textcolor{black}{#1}}

\definecolor{codebg}{rgb}{0.97,0.97,0.94}
\definecolor{codekeyword}{rgb}{0.00,0.30,0.70}
\definecolor{codestring}{rgb}{0.65,0.15,0.15}
\definecolor{codecomment}{rgb}{0.25,0.55,0.30}
\definecolor{codetype}{rgb}{0.55,0.20,0.55}
\definecolor{codenumber}{rgb}{0.50,0.50,0.50}
\definecolor{codeframe}{rgb}{0.80,0.80,0.80}

\lstdefinestyle{cstyle}{
  language=C,
  morekeywords={struct},
  emph={int,char,void,float,double,unsigned,size_t},
  emphstyle=\color{codetype}\bfseries,
}
\lstdefinelanguage{riscv}{
  morekeywords={
    ld,sd,lw,sw,lb,sb,lh,sh,lui,auipc,
    add,addi,sub,and,andi,or,ori,xor,xori,sll,slli,srl,srli,sra,srai,
    beq,bne,blt,bge,bltu,bgeu,
    jal,jalr,j,jr,ret,call,
    mv,li,la,nop,
    bld,brl
  },
  morekeywords=[2]{
    zero,ra,sp,gp,tp,fp,
    t0,t1,t2,t3,t4,t5,t6,
    a0,a1,a2,a3,a4,a5,a6,a7,
    s0,s1,s2,s3,s4,s5,s6,s7,s8,s9,s10,s11,
    rs1,rs2,rd
  },
  morecomment=[l]{\#},
  sensitive=true,
}
\lstdefinestyle{asmstyle}{
  language=riscv,
  keywordstyle=\color{codekeyword}\bfseries,
  keywordstyle=[2]\color{codetype},   
  commentstyle=\color{codecomment}\itshape,
}

\begin{document}
\title{Branch Landing: Bloom Filter-Based Source Authorization for Forward-Edge CFI on RISC-V}
\titlerunning{Branch Landing}
%
\author{You Wu\inst{1}\orcidID{0000-0002-1410-3196} \and
Peter Beerel\inst{1}\orcidID{0000-0002-8283-0168}}
\authorrunning{Y. Wu et al.}
%
\institute{University of Southern California, Los Angeles CA 90057, USA 
\email{\{youwu, pabeerel\}@usc.edu}}
\maketitle              
\begin{abstract}
Jump-Oriented Programming (JOP) attacks exploit indirect control transfers to bypass backward-edge defenses, yet existing forward-edge CFI mechanisms lack precise source-domain authorization: type-based CFI admits all same-signature callers, while tag-based hardware CFI is limited by fixed-width register storage that caps the number of simultaneously authorized sources.
We propose Branch Landing (BRL), a landing-based forward-edge CFI framework for RISC-V that replaces fixed-capacity checks with Bloom filter membership queries.
Two lightweight ISA extensions, \texttt{bld} and \texttt{brl}, propagate a source Section Identifier (SID) through a dedicated \texttt{BRState} register and validate it at each landing site with fixed-probe latency that is independent of the number of authorized sources under a chosen filter configuration.
Section granularity is configurable, supporting policies from type-based to CFG-derived authorization within a single mechanism.
We implement Branch Landing in the LLVM RISC-V backend and evaluate it on 81 BEEBS benchmarks under two representative policy configurations: a function-level, type-based policy and a basic-block-level, CFG-derived policy.
Under a 3-cycle brl latency model, the two configurations incur average runtime overheads of only 0.210\% and 0.421\%, with mean code size growth of 0.46\% and 0.52\% respectively.
The CFG-derived policy reduces the average equivalence class size by 32.5\% compared to the type-based policy, and all evaluated executions complete without BRL enforcement failures.
\end{abstract}

\section{Introduction}
\label{sec:intro}

Memory corruption vulnerabilities continue to pose a fundamental threat to system security.
Even in the presence of Data Execution Prevention (DEP)~\cite{dep2003} and Address Space Layout Randomization (ASLR)~\cite{pax2003aslr, shacham2004aslr}, attackers can hijack control flow through code-reuse attacks~\cite{roemer2012rop, bletsch2011jop, schuster2015coop}.
While shadow stacks~\cite{burow2019sok} and backward-edge protections have significantly reduced practical Return-Oriented Programming (ROP) exploitability, attackers increasingly pivot to Jump-Oriented Programming (JOP)~\cite{bletsch2011jop, checkoway2010rop_without_ret}, which chains gadgets exclusively through forward-edge indirect transfers such as \texttt{jalr} and \texttt{jr} in RISC-V~\cite{riscv-isa}.
Since JOP bypasses backward-edge defenses entirely, dedicated forward-edge CFI mechanisms are essential~\cite{abadi2009cfi, burow2017cfi_survey}.
 
Existing JOP defenses fundamentally leave a \emph{source-authorization gap}, as they fail to precisely bind indirect branches to their legitimate origins.
Coarse landing-based mechanisms, such as Intel CET~\cite{intel-cet, shanbhogue2019cet} and ARM BTI~\cite{arm-bti}, verify that an indirect branch reaches a valid destination but impose no constraint on \emph{which source} may initiate the transfer.
Type-based CFI mechanisms, such as FineIBT~\cite{fineibt} and Clang CFI~\cite{llvm-cfi, tice2014vtable}, narrow the target set through function-signature matching, but all callers sharing the same type pass the check indiscriminately.
Tag-based hardware mechanisms, such as Bratter~\cite{bratter}, support richer context-sensitive policies via per-source tag values, but their fixed-width register storage imposes a hard ceiling on the number of sources that can be simultaneously authorized for a single target. Consequently, when legitimate callers exceed the available tag slots, the remaining callers are left without an enforceable policy.
 
In this paper, we propose \textbf{Branch Landing (BRL)}, a landing-based forward-edge CFI framework for RISC-V that shifts the hardware enforcement paradigm from rigid one-to-one equality checks to flexible, set-based membership queries. 
By employing Bloom filters~\cite{bloom1970filter} to represent authorized caller sets, BRL effectively decouples authorization capacity from fixed-width register constraints.
To realize this paradigm shift at the architectural level, we design two novel, lightweight ISA extensions: \texttt{bld}~(Branch Landing Descriptor) and \texttt{brl}~(Branch Landing Verification). 
Prior to an indirect branch, our \texttt{bld} instruction securely propagates a source Section Identifier (SID) into a dedicated architectural register, \texttt{BRState}. 
Subsequently, the \texttt{brl} instruction operates at the valid landing site to evaluate the condition $\text{SID}_{\text{src}} \in \textit{AllowedSources}(T)$ with a fixed number of hash probes independent of $\textit{AllowedSources}(T)$, immediately triggering a control-flow protection fault if the source SID falls outside the authorized set.

Two key properties distinguish Branch Landing from prior mechanisms. 
First, it provides \emph{scalable source-domain authorization}: unlike type-based CFI, which reduces all callers to a single equivalence class, and unlike tag-based CFI, which is structurally limited by register slot count, BRL leverages its underlying Bloom filter to encode large authorized-source sets per target with fixed-probe verification and no general-purpose register pressure.
Second, our instruction design is inherently \emph{policy-agnostic}: the newly introduced \texttt{bld}/\texttt{brl} hardware infrastructure remains static, yet it can instantiate type-based, CFG-derived~\cite{niu2015mcfi, zhang2013ccfir}, or other authorization models by varying only how $\textit{AllowedSources}(T)$ is constructed at compile time. 
Table~\ref{tab:comparison} summarizes these design axes across representative CFI mechanisms.
 
\begin{table}[t]
\centering
\caption{Comparison of forward-edge CFI mechanisms.}
\label{tab:comparison}
\resizebox{\columnwidth}{!}{%
\begin{tabular}{lcccc}
\toprule
Mechanism & Source-aware & Auth.\ model & Scalable & Policy-agnostic \\
\midrule
CET/BTI~\cite{intel-cet,arm-bti}              & \ding{55}    & Target-only   & \checkmark & \ding{55} \\
FineIBT~\cite{fineibt}              & Partial      & Type equality & \checkmark & \ding{55} \\
Bratter~\cite{bratter}              & \checkmark   & Tag equality  & \ding{55}  & Partial \\
\textbf{Branch Landing (BRL)} & \checkmark & Membership  & \checkmark & \checkmark \\
\bottomrule
\end{tabular}}
\end{table}

We implement Branch Landing in the LLVM RISC-V backend and evaluate
it on the 81 BEEBS benchmarks~\cite{pallister2013beebs} under two policy
configurations: \emph{BRL-Func} (function-level SID with type-based
authorization) and \emph{BRL-CFG} (basic-block-level SID with
CFG-derived authorization). 
Runtime overhead is measured on the Spike RISC-V ISA simulator using a weighted-cycle model under three \texttt{brl} latency points: 3, 5, and 10 cycles, with bld modeled as a 1-cycle state update, following the methodology of Bratter~\cite{bratter}. The \texttt{brl3}, \texttt{brl5}, and \texttt{brl10} represent optimistic, realistic, and conservative hardware implementations, respectively. A detailed cycle-by-cycle justification is given in Section~\ref{sec:discussion-latency}.
Under the \texttt{brl3} model,
BRL-Func and BRL-CFG incur mean runtime overheads of
0.210\% and 0.421\% respectively, 
more than 14$\times$ lower than Bratter's reported 5.99\% for combined function-signature
and branch regulation enforcement under the same weighted-cycle style
of evaluation.
Even under the conservative
\texttt{brl10} model, the mean overhead remains below 1.3\%.
This comparison should
be interpreted as indicative rather than identical, since the protected edge sets
and hardware assumptions differ across the two systems.
Code size overhead remains modest
at 0.46\% (BRL-Func) and 0.52\% (BRL-CFG)  mean, less than
half of Bratter\_Both's 1.20\%, while BRL-CFG simultaneously reduces
the average equivalence class size by 32.5\% relative to BRL-Func.
All 81 benchmarks execute with zero false positives across both
configurations.

In summary, this paper makes the following contributions:
\begin{itemize}
    \item We formalize the source-authorization gap in existing type-based and tag-based forward-edge CFI mechanisms and propose Branch Landing, a RISC-V CFI framework that enforces source-domain authorization via Bloom filter membership queries through two lightweight ISA extensions.
    \item We introduce a configurable section-granularity model that instantiates multiple CFI policies, ranging from type-based to CFG-derived, within a single policy-agnostic enforcement mechanism.
    \item We implement Branch Landing as an LLVM RISC-V backend extension and a Spike ISA simulator prototype, demonstrating on the BEEBS benchmark suite that it achieves substantially lower runtime 
    and code size overhead than prior tag-based hardware CFI while providing finer-grained source authorization and tighter equivalence classes. 
\end{itemize}
 
The remainder of this paper is organized as follows.
Section~\ref{sec:background} reviews code-reuse attacks and related CFI mechanisms.
Section~\ref{sec:design} presents the design of Branch Landing.
Section~\ref{sec:eval} evaluates Branch Landing on the BEEBS benchmark suite.
Section~\ref{sec:discussion} discusses design considerations and potential extensions.
Section~\ref{sec:conclusion} concludes.

\section{Background and Related Work}
\label{sec:background}
This section reviews prior work on control-flow integrity relevant to
Branch Landing, organized along the design axes that motivate our
approach: attack model, enforcement granularity, and authorization model.

\subsection{Jump-Oriented Programming on RISC-V}
\label{sec:bg-jop}

Memory corruption vulnerabilities, such as buffer overflows, use-after-free errors, and type confusion, allow attackers to overwrite control-sensitive data including return addresses, function pointers, and jump table indices.
Return-Oriented Programming (ROP)~\cite{roemer2012rop} chains short instruction sequences ending in \texttt{ret} to construct arbitrary computations; shadow stacks~\cite{burow2019sok} have substantially mitigated this vector.
Jump-Oriented Programming (JOP)~\cite{bletsch2011jop} avoids return instructions entirely, instead sequencing gadgets through indirect control transfers. Because JOP uses only forward-edge transfers, it
bypasses shadow stacks and other backward-edge defenses, motivating
dedicated forward-edge CFI mechanisms.

On RISC-V, the \texttt{jalr} instruction is the primary JOP primitive.
A typical attack proceeds as follows. Consider a program that stores a
function pointer in memory and later invokes it through an indirect
call:

\begin{lstlisting}[style=cstyle,
                   caption={Vulnerable indirect call site.},
                   label={lst:jop-c}]
struct handler {
    char buf[64];
    void (*fn)(int);   // function pointer
};

void dispatch(struct handler *h, int arg) {
    h->fn(arg);        // compiles to jalr ra, rs1, 0
}
\end{lstlisting}

A buffer overflow into \texttt{h->buf} lets the attacker overwrite
\texttt{h->fn} with the address of a chosen gadget. When
\texttt{dispatch} executes the indirect call, control transfers to that
gadget instead of the intended callee. A useful JOP gadget on RISC-V
takes the form:

\begin{lstlisting}[style=asmstyle,
                   caption={A representative RISC-V JOP gadget.},
                   label={lst:jop-asm}]
# Load next target from an attacker-controlled
# dispatcher table and jump to it.
ld    t0, 0(a0)      # load next gadget address
addi  a0, a0, 8      # advance dispatcher pointer
jalr  zero, t0, 0    # indirect jump to next gadget
\end{lstlisting}

By chaining such gadgets through a dispatcher table, the attacker
constructs Turing-complete computation. Crucially, every transfer in the
chain is a syntactically legitimate \texttt{jalr} to a valid
instruction address, so coarse landing-based defenses that only check whether the destination
is a marked landing site fail to detect the attack: the gadget entry
points are themselves valid landing sites in the original program.
This observation motivates \emph{source-aware} CFI, in which each
indirect transfer must be authorized not only by its destination but
also by its origin.

\begin{figure*}[t]
  \centering
  \includegraphics[width=\linewidth]{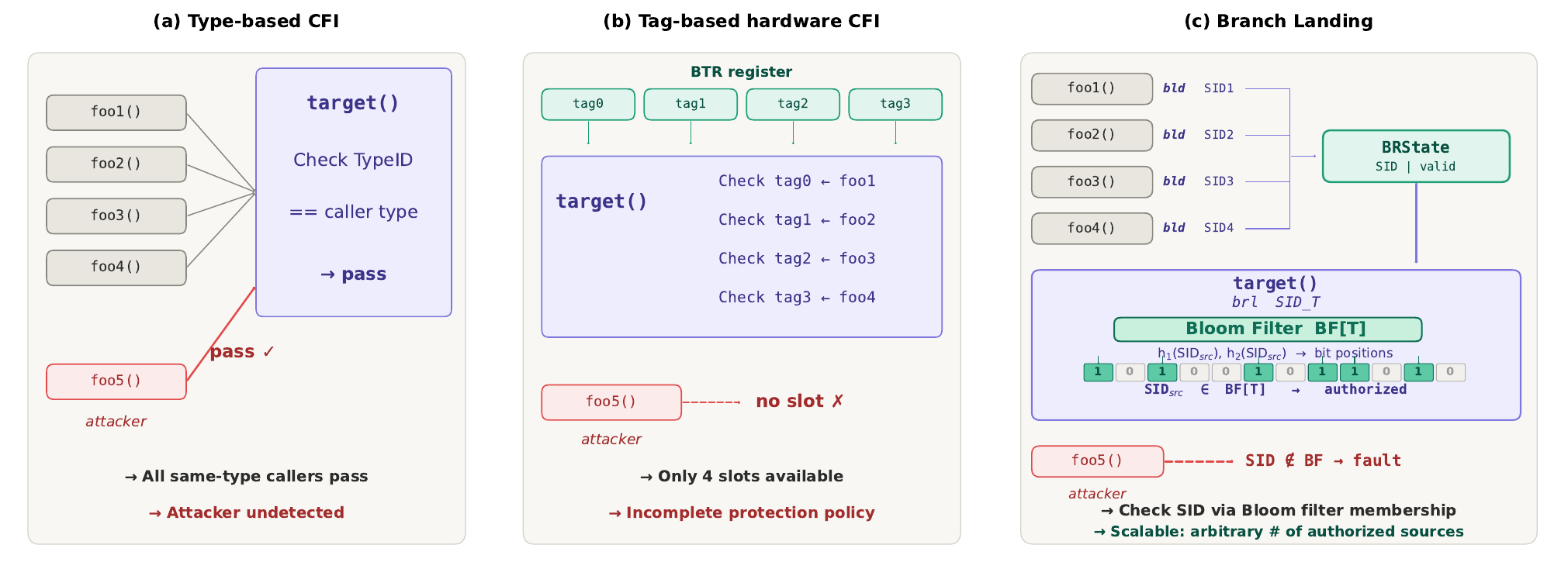}
  \caption{Existing and proposed forward-edge CFI defenses.
           (a)~Type-based CFI (FineIBT): all callers sharing the same
           function type pass the check, including attacker-controlled
           \texttt{foo5}.
           (b)~Tag-based hardware CFI (Bratter): the 4-slot BTR is
           exhausted by \texttt{foo1}–\texttt{foo4}; no slot is
           available to express a policy for \texttt{foo5}.
           (c)~Branch Landing: source SIDs are checked via Bloom filter
           membership, supporting an arbitrary number of authorized
           callers with fixed-probe verification independent of set size.}
  \label{fig:motivation}
\end{figure*}

\subsection{Coarse Landing-Based CFI}
Intel CET IBT~\cite{intel-cet,shanbhogue2019cet}, ARM BTI~\cite{arm-bti}, and
recent RISC-V proposals~\cite{christoulakis2016hcfi,riscv-cfi,manoni2026cva6} require
indirect branches to land on designated marker instructions
(\texttt{endbr}, \texttt{bti}, or equivalent).  These mechanisms
are \emph{target-centric}: they verify that the destination is a
valid landing site but impose no constraint on which source may
initiate the transfer, admitting JOP chains among the full set of
marked sites.

\subsection{Type-Based CFI}
FineIBT~\cite{fineibt} and Clang CFI~\cite{llvm-cfi,tice2014vtable} restrict
indirect calls to targets with compatible function signatures.
Authorization remains a single equality check: all call sites
sharing the same type reach all matching targets, regardless of
structural caller--callee relationships.

ECCut~\cite{eccut} refines LLVM-CFI equivalence classes through
complete field sensitivity, achieving 94.8\% average EC reduction
on SPEC CPU; HPCFI~\cite{kasten2024hpcfi} integrates multiple static
analyses for similar gains.  Both narrow the type-based target set
but remain equality-based and do not provide explicit source-domain
authorization.  Their analysis refinements are complementary to
BRL and could serve as policy inputs to construct tighter
\texttt{AllowedSources}$(T)$ sets.

\subsection{Tag-Based Hardware CFI}
 
Tag-based mechanisms associate each source with a runtime tag value
and verify it at the landing site, supporting finer-grained policies
than type equality alone.
 
Bratter~\cite{bratter} introduces a Branch Tag Register (BTR), a
32-bit CSR partitioned into four independent 8-bit slots
(\texttt{tag0}--\texttt{tag3}). Two dedicated instructions,
\texttt{sbtag} and \texttt{cbtag}, write and verify individual slots
before and after indirect transfers, respectively. This allows Bratter
to enforce context-sensitive policies that would require trampolines or
function duplication under single-tag mechanisms. Bratter reports
average execution time overheads of 0.43\% for function-signature
enforcement and 5.99\% when combined with branch regulation on the
BEEBS benchmark suite.
 
However, Bratter's enforcement model remains equality-based: each slot
stores a fixed tag value and the callee checks for an exact match.
When multiple legitimate callers with different contexts must reach the
same target, they must either share a tag or occupy separate slots. The four-slot BTR imposes a hard
ceiling: as shown in Figure~\ref{fig:motivation}(b), a program with five
or more distinct source domains targeting the same function exhausts
the register, leaving the remaining callers without an enforceable
policy. This is a structural limitation of fixed-width tag storage,
not an artifact of any particular implementation.

\section{Design}
\label{sec:design}

Branch Landing enforces forward-edge CFI through two lightweight ISA
extensions, \texttt{bld} and \texttt{brl}, combined with per-target
Bloom filter metadata. This section describes the threat model, the
core enforcement mechanism, the supported policy instantiations, and
the granularity trade-offs.
 
\subsection{Threat Model}
\label{sec:threat}

Branch Landing targets forward-edge control-flow hijacking attacks,
including Jump-Oriented Programming (JOP), function pointer corruption,
virtual call hijacking, and dispatcher-based gadget chaining. The
attacker is assumed to possess arbitrary memory write capabilities---sufficient to corrupt function pointers, virtual table entries, or jump table indices, but
is unable to modify executable code or Branch Landing metadata, which
are protected as read-only by the operating system.  We further assume that the
attacker cannot tamper with the \texttt{BRState} register, 
which is not memory-mapped and is updated architecturally only by committed bld/brl instructions rather than by ordinary stores
or generic user-level CSR writes (Section~\ref{sec:brstate}). 
Under this model, the defense guarantees that any indirect control
transfer must originate from a source domain authorized to reach the
target landing site; unauthorized transfers trigger a processor
exception.

We explicitly place microarchitectural side-channel attackers outside
our threat model. Branch Landing is an architectural CFI mechanism
and inherits the same scoping convention as prior hardware-assisted
CFI work, including FineIBT~\cite{fineibt}, Bratter~\cite{bratter},
and Intel CET~\cite{intel-cet}, all of which target architectural control
flow rather than microarchitectural leakage. Speculative execution,
cache timing, and other microarchitectural channels require orthogonal
defenses (e.g., speculative load hardening, cache partitioning) that
can be deployed alongside BRL. We analyze the residual side-channel
attack surface that BRL specifically introduces, and discuss potential
hardening directions, in Section~\ref{sec:security-analysis}.

\subsection{ISA Extensions}
\label{sec:isa}
Branch Landing introduces two new instructions to the RISC-V ISA.
Both are encoded as I-type instructions within a reserved custom
opcode space, ensuring backward compatibility with existing RISC-V
implementations: on legacy hardware that does not recognize
\texttt{bld}/\texttt{brl}, the instructions can be treated as
illegal-instruction faults, providing a fail-closed default.
 
\paragraph{\texttt{bld} (Branch Landing Descriptor).}
The \texttt{bld} instruction is inserted by the compiler immediately
before each indirect control transfer (\texttt{jalr} or \texttt{jr}).
It takes an assembler-supplied source Section Identifier (SID) operand
and performs two operations atomically:
(1)~it writes the SID into \texttt{BRState.sid}, and
(2)~it sets the validity bit \texttt{BRState.valid} $\leftarrow 1$.
The instruction executes in a single cycle, as it only writes an
immediate value into a dedicated CSR without requiring memory access
or ALU computation. 

\begin{center}
\texttt{bld \; SID\_src} \\
\texttt{jalr \; x0, rs1, 0}
\end{center}

\texttt{brl (Branch Landing Verification).}
The \texttt{brl} instruction is inserted at the entry of each valid
indirect branch target. 
It takes a target Section Identifier
($SID_T$) operand that uniquely identifies the landing site within the program.
When execution reaches
a landing site, \texttt{brl} performs the following sequence:
\begin{enumerate}
    \item \textbf{Validity check.} If $\texttt{BRState.valid} = 0$,
    the site was reached without a preceding \texttt{bld}, indicating
    an attacker-controlled jump that bypassed the instrumented call
    sequence. The processor raises a control-flow protection fault
    immediately.
    \item \textbf{Filter resolution.} \texttt{brl} uses its target 
    SID operand to index a global filter descriptor table stored in
    \texttt{.rodata}, retrieving the base address and bit-width of
    the per-target Bloom filter $BF[SID_T]$. The descriptor table
    layout is fixed at link time and protected as read-only.
    \item \textbf{Bloom filter lookup.} \texttt{brl} retrieves
    $SID_{src}$ from \texttt{BRState.sid} and evaluates the
    membership query $SID_{src} \in BF[SID_T]$ by hashing $SID_{src}$
    through $k$ hash functions and checking the corresponding bits
    in $BF[SID_T]$.
    \item \textbf{Authorization decision.} If all $k$ bits are set,
    the transfer is authorized: execution proceeds normally and
    \texttt{BRState.valid} is cleared to 0, ensuring that
    authorization state cannot be reused across multiple transfers.
    If any bit is clear, $SID_{src}$ is not a member of the
    authorized set and a control-flow protection fault is raised.
\end{enumerate}

Under the latency models used in our evaluation, \texttt{brl} performs
membership checking with a fixed number of hash probes, in time independent of the number of authorized source domains for the target.
This timing assumes a
hardware-optimized implementation in which the filter descriptor and
filter words hit L1 D-cache and the two hash functions execute in
parallel; we revisit these assumptions and their hardware implications
in Section~\ref{sec:discussion-latency}.

\subsection{BRState Register}
\label{sec:brstate}
 
Branch Landing introduces a dedicated architectural register,
\texttt{BRState}, implemented as a Control and Status Register (CSR)
whose fields are not directly
writable by application memory operations or generic user-level CSR writes.
Application code updates this state only through the new bld/brl instructions,
while privileged software may save, restore, or clear it as part of architectural
context management. \texttt{BRState} maintains two fields:
 
\begin{itemize}
  \item \texttt{BRState.valid} (1 bit): indicates whether a pending
        source authorization is active.  Set to~$1$ by \texttt{bld};
        cleared to~$0$ by \texttt{brl} upon completion of the
        landing-site check.
  \item \texttt{BRState.sid} (31 bits): holds the source Section
        Identifier propagated by the most recent \texttt{bld}.  A
        31-bit field supports up to $2^{31}$ distinct SIDs, sufficient
        for basic-block-level granularity even in large programs.
\end{itemize}
 
\texttt{BRState} is a \emph{single-use} register: the validity bit
is cleared by every \texttt{brl} execution, ensuring that
authorization state from one transfer cannot be carried over to a
subsequent transfer.  This single-use invariant is critical to
security: without it, an attacker who reaches a \texttt{brl} site via
a legitimate transfer could retain the authorized \texttt{BRState} and
redirect a second transfer to an unauthorized target.

\paragraph{Interaction with Exceptions and Context Switches.}
When a trap or context switch occurs between a \texttt{bld} and its
corresponding \texttt{brl}, the operating system must save and restore
\texttt{BRState} as part of the thread context, similar to the
treatment of other privileged CSRs.  If \texttt{BRState} is not
preserved, the subsequent \texttt{brl} will observe
\texttt{BRState.valid} $= 0$ and raise a fault. This
fail-closed behavior prevents authorization state from leaking across
execution contexts.  An alternative design choice is to
unconditionally clear \texttt{BRState} on every trap entry; this
sacrifices the rare case where an interrupt occurs within the
\texttt{bld}--\texttt{jalr}--\texttt{brl} window unless the trap handler restarts execution before the consumed bld but simplifies the OS context-switch path.

\subsection{Section ID Assignment and Granularity}
\label{sec:sid-granularity}
 
Branch Landing organizes program code into protection domains called
\emph{sections}, each assigned a unique Section Identifier (SID) at
compile time.  The granularity of section assignment is configurable
along a spectrum:
 
\begin{itemize}
  \item \textbf{Module granularity:} all functions within a compilation
        unit share a single SID.  This minimizes metadata size but
        admits any caller within the same module, providing the weakest
        isolation.
  \item \textbf{Function granularity:} each function receives a
        distinct SID.  This is the natural match for type-based
        authorization policies and is used in the BRL-Func configuration.
  \item \textbf{Basic-block granularity:} each basic block is assigned
        an independent SID, so that different call sites within the same
        function carry distinct source identities.  This is the
        granularity used in BRL-CFG and prevents intra-function
        authorization reuse.
  \item \textbf{User-defined groupings:} the compiler accepts
        annotations that group basic blocks into custom protection
        domains, enabling application-specific policies such as isolating
        privilege-boundary code from general-purpose code within the
        same function.
\end{itemize}
 
The choice of granularity affects three dimensions simultaneously.
Finer granularity strengthens isolation. An attacker who compromises
one call site within a function cannot reuse its SID to reach targets
authorized only for a different call site, but increases the number
of distinct SID values in the program, which in turn affects Bloom
filter sizing.  Coarser granularity reduces the SID namespace and
shrinks per-target Bloom filters, but admits more source domains per
authorization set.  Cross-compilation-unit SID assignment is
straightforward at function granularity (SIDs can be assigned per
compilation unit and reconciled at link time) but requires either
link-time optimization (LTO) or a link-time metadata merge pass at
basic-block granularity, where the SID namespace grows substantially.
Our current prototype assumes whole-program compilation; extending the
toolchain to support incremental compilation with basic-block SIDs is
a practical engineering challenge that does not affect the hardware
mechanism.

\subsection{Bloom Filter Authorization Metadata}
\label{sec:bloom}
 
For each landing site $T$, the compiler constructs the authorization
set $\textit{AllowedSources}(T)$ from the selected CFI policy and
encodes it into a Bloom filter stored in a read-only metadata section
(\texttt{.rodata}).  A Bloom filter is a space-efficient probabilistic
data structure that supports membership queries with a fixed number of hash probes,
a tunable false positive rate and zero false negatives~\cite{bloom1970filter}.
 
\paragraph{Filter Parameters.}
Each per-target filter is characterized by three parameters: $m$ (the
number of bits), $k$ (the number of independent hash functions), and
$n$ (the number of authorized SIDs to encode).  The theoretical false
positive rate is approximately:
\begin{equation}
  p_{\text{fp}} \;\approx\; \bigl(1 - e^{-kn/m}\bigr)^{k}
  \label{eq:fp}
\end{equation}
A lower $p_{\text{fp}}$ tightens the security guarantee, reducing the
probability that an unauthorized SID is spuriously admitted at the
cost of a larger $m$.  In our evaluation, we use a fixed filter size
that provides $p_{\text{fp}} < 10^{-3}$ across all benchmarks.  For
programs with very large $\textit{AllowedSources}$ sets (e.g.,
event-driven frameworks with hundreds of registered callbacks),
per-target adaptive sizing can maintain a uniform false positive bound
without inflating metadata for simple targets.
 
\paragraph{Hash Function Design.}
The $k$ hash functions must be fast, independent, and deterministic.
We employ a double-hashing scheme~\cite{kirsch2006less}: given two
base hash functions $h_1$ and $h_2$, the $i$-th hash is computed as
$h_i(\text{SID}) = (h_1(\text{SID}) + i \cdot h_2(\text{SID}))
\bmod m$ for $i = 0, \ldots, k{-}1$.  This reduces the number of
independent hash computations to two while preserving the theoretical
false positive bound, and maps naturally to a two-cycle hardware
implementation within the \texttt{brl} instruction's execution.
 
\paragraph{Metadata Layout.}
Per-target Bloom filter bit arrays are stored contiguously in a
dedicated read-only section, indexed by target SID. A global
\emph{filter descriptor table} maps each $SID_T$ to a tuple
$(\mathit{base}, m)$, where $\mathit{base}$ points to the start of
$BF[SID_T]$ within the bit-array region and $m$ is the filter
bit-width. At runtime, the immediate operand carried by each
\texttt{brl} instruction directly indexes this descriptor table,
yielding the metadata required to perform the membership check
described in Section~\ref{sec:isa}. Both the descriptor table and
the filter bit arrays reside in read-only memory protected by the
operating system's page permissions; an attacker with arbitrary
write capabilities cannot modify the authorization sets without
triggering a page fault, consistent with the threat model defined
in Section~\ref{sec:threat}.
 
\paragraph{Interaction with SID Granularity.}
The Bloom filter size and SID granularity interact in a
counterintuitive way: finer SID granularity tends to \emph{shrink}
each target's authorized set (because the CFG analysis distributes
callers over more distinct SIDs), which in turn \emph{reduces} the
required filter size per target.  Conversely, coarser granularity
collapses callers into fewer SIDs, yielding trivially smaller
authorized sets but with weaker security.  The net metadata cost
therefore depends on the program structure: callback-heavy programs
benefit most from fine SIDs and small filters, while programs
dominated by a few highly polymorphic call sites may see diminishing
returns from finer granularity.

\subsection{Policy Instantiations}
 
Branch Landing decouples policy generation from runtime enforcement.
The same \texttt{bld}/\texttt{brl} mechanism can enforce different CFI
policies by varying how $\mathrm{AllowedSources}(T)$ is constructed
at compile time. We evaluate two representative configurations.
 
\paragraph{BRL-Func (Function-Level Granularity).}
Each function is assigned a unique SID and authorization sets are
derived from function type signatures: a target $T$ admits any source
whose call site declares a compatible type. This configuration is comparable in
policy strength to FineIBT~\cite{fineibt} and serves as a performance
baseline. Unlike
FineIBT, which performs a single type-equality check, BRL-Func uses a
Bloom filter membership query, naturally handling the case where multiple
call sites of the same type are authorized without requiring a shared tag.
 
\paragraph{BRL-CFG (Basic-Block-Level Granularity).}
Authorization sets are derived from an interprocedural control-flow
graph. SIDs are assigned at basic-block granularity---one SID per basic block---so
that each call site within a function carries a distinct source identity.
The compiler projects CFG edges onto
protection domain pairs, producing a refined $\mathrm{AllowedSources}(T)$
that admits only those domains from which a direct CFG path to $T$
exists. 
This combination of finer-grained
SID assignment and CFG-derived policy prevents an attacker from reusing
authorization obtained at one call site to reach a different target through
another call site within the same function, approaching the precision of
whole-program CFG-based defenses while retaining the compatibility benefits
of landing-based enforcement.
 
The two configurations represent different points in the
security--overhead trade-off space, which we quantify in
Section~\ref{sec:eval}.

\subsection{Security Analysis}
\label{sec:security-analysis}

We analyze the security guarantees of Branch Landing under the threat
model defined in Section~\ref{sec:threat}.

\subsubsection{Security Invariants}

Branch Landing's enforcement guarantees rest on three invariants
jointly maintained by the hardware mechanism, the compiler
instrumentation, and the operating system's memory protection.

\paragraph{Invariant 1 (Authorized Origin).}
Every successful execution of \texttt{brl} at landing site~$T$
implies that a preceding \texttt{bld} instruction was committed on
the same control-flow path, and that the source SID it propagated
satisfies $SID_{src} \in BF[SID_T]$.

\paragraph{Invariant 2 (Single-Use Authorization).}
The \texttt{BRState.valid} bit is monotonically consumed: any
\texttt{brl} that observes $\texttt{valid}=1$ clears it before
commit, and any \texttt{brl} that observes $\texttt{valid}=0$
raises a control-flow protection fault.

\paragraph{Invariant 3 (Metadata and Code Integrity).}
The filter descriptor table, all per-target Bloom filter bit arrays,
and the program text containing \texttt{bld} immediates reside in
read-only memory protected by OS page permissions. \texttt{BRState}
is writable only through committed \texttt{bld} instructions.

\subsubsection{Defense Against Forward-Edge Attacks}

Invariant~1 blocks \emph{JOP gadget chaining} and \emph{SID forgery
via memory corruption}: a corrupted indirect branch either lacks a
preceding \texttt{bld} altogether or carries a compile-time-fixed
SID that is not a member of the target's authorized set, causing the
subsequent \texttt{brl} to fault.  Because \texttt{BRState} is a
privileged CSR whose \texttt{sid} field can only be updated by
executing an existing \texttt{bld} instruction in read-only program
text (Invariant~3), the attacker cannot inject arbitrary SID values.
Invariant~2 covers two additional cases: an attacker who jumps
directly to a \texttt{brl} site without a preceding \texttt{bld}
observes $\texttt{valid}=0$ and faults immediately (\emph{bld
bypass}), and authorization state consumed by one \texttt{brl}
cannot be replayed at a second landing site (\emph{BRState replay}).
Finally, Invariant~3 prevents \emph{filter or descriptor tampering}:
both reside in \texttt{.rodata} and any write triggers a page fault.

\subsubsection{Bloom Filter False Positives}

An attacker cannot inject arbitrary SIDs into \texttt{BRState}; the
only architectural path is to execute an existing \texttt{bld} whose
immediate is fixed at compile time (Invariant~3).  Let~$\mathcal{S}$
denote the set of distinct SIDs in the binary.  The expected number
of spuriously authorized SIDs at any target~$T$ is bounded by
$|\mathcal{S}| \cdot p_{\mathit{fp}}$; with
$|\mathcal{S}| \sim 10^{3}$ and $p_{\mathit{fp}} < 10^{-3}$ in our
benchmarks, fewer than one false authorization is expected per target.
Exploiting even a spurious SID further requires a feasible path to
the carrier \texttt{bld} \emph{and} a hijackable indirect branch
to~$T$, making the practical attack surface substantially smaller
than the raw~$p_{\mathit{fp}}$.  We measured zero false positives
across all 81~benchmarks under both configurations
(Section~\ref{sec:eval}).

\subsubsection{Limitations}

BRL protects forward edges only; backward-edge attacks require a
complementary shadow stack.  An adversary who legitimately reaches a
landing site and exploits code beyond it is not detected, a
limitation shared by all landing-based
CFI~\cite{carlini2015cfb}.  Without LTO, independently compiled
units may emit colliding SIDs; a link-time merge pass is required.
BRL introduces filter-access and timing side channels analogous to
those in prior hardware
CFI~\cite{fineibt,bratter,intel-cet}; mitigating them requires orthogonal
microarchitectural defenses and is left to future work.  Data-only
attacks~\cite{hu2016dop} are outside the scope of any CFI mechanism.

\section{Evaluation}
\label{sec:eval}
We implemented Branch Landing in the LLVM RISC-V backend,
adding compiler instrumentation for \texttt{bld} insertion at
indirect branch sources and \texttt{brl} insertion at valid landing
sites, together with Bloom filter metadata generation for per-target
authorization.  Runtime evaluation is performed on the Spike RISC-V
ISA simulator extended with \texttt{bld}/\texttt{brl} support.
 
\subsection{Evaluation Configurations}
We evaluate three configurations:
\begin{itemize}
    \item \textbf{Baseline}: an uninstrumented build without
          control-flow protection.
    \item \textbf{BRL-Func}: each function receives a unique SID;
          authorization sets are derived from function type signatures,
          mirroring the policy strength of type-based CFI mechanisms
          such as FineIBT~\cite{fineibt}.
    \item \textbf{BRL-CFG}: SIDs are assigned at basic-block
          granularity (one SID per basic block); authorization sets
          are derived from interprocedural control-flow graph analysis,
          admitting only structurally reachable callers.
\end{itemize}
BRL-Func isolates the cost of Bloom filter enforcement under a
coarse, type-equivalent policy; BRL-CFG represents the strictest
instantiation, combining finer-grained SID assignment with a more
precise authorization policy.

We use the BEEBS (Bristol/Embecosm Embedded Benchmark Suite)~\cite{pallister2013beebs},
a collection of 81 small C programs designed for bare-metal execution
and commonly used to evaluate compiler optimizations and ISA
extensions.  All benchmarks are compiled with the same LLVM toolchain
at optimization level \texttt{-O2} under each configuration. 
We report the arithmetic mean across all benchmarks, consistent with the methodology used by Bratter to enable direct comparison.
 
We report three classes of metrics:
(1)~\emph{code size}: static \texttt{.text} section growth and CFI
    instruction density (ratio of dynamically executed
    \texttt{brl}/\texttt{bld} instructions to total instructions);
(2)~\emph{performance}: weighted execution time overhead under three
    latency assumptions---\texttt{brl3} (3-cycle \texttt{brl}), \texttt{brl5} (5-cycle \texttt{brl}) and \texttt{brl10} (10-cycle \texttt{brl}), all paired with a 1-cycle bld.
(3)~\emph{security}: equivalence class (EC) size and the number of
    protected targets under each policy;

\subsection{Code Size Overhead}
\label{sec:eval-codesize}

We measure code size overhead introduced by Branch Landing by
comparing the \texttt{.text} section size of instrumented binaries
against an uninstrumented baseline. This ensures that any difference in \texttt{.text}
size is attributable solely to the inserted \texttt{BRL}/\texttt{BLD}
instructions and Bloom filter lookup sequences.

Table~\ref{tab:codesize} reports the results for selected benchmarks
and aggregate statistics across all 81 BEEBS benchmarks.  The overhead
is consistently small: on average \textbf{0.46\%} for
BRL-Func and \textbf{0.52\%} for BRL-CFG.   

Under BRL-CFG, most benchmarks exhibit overhead comparable to or
lower than BRL-Func, because the pointer analysis prunes spurious
targets and produces smaller Bloom filters.  However, a few benchmarks
show elevated CFG overhead.  The most notable is \texttt{cover}
(7.85\%), where the pointer analysis resolves 181 indirect call
targets---far more than the 2 identified by type matching---leading
to additional Bloom filter entries.  Similarly, \texttt{trio-snprintf}
(2.30\%) and \texttt{trio-sscanf} (1.88\%) contain rich
format-string dispatch logic that the CFG analysis resolves more
completely.  These cases represent a deliberate security--size
tradeoff: the CFG policy protects a strictly larger set of indirect
transfers at the cost of modestly larger instrumentation metadata.

The low code size overhead stems from the Bloom filter encoding:
each target's allowed-source set is stored as a compact bit array,
so a single \texttt{BRL} instruction suffices at each protected
entry point regardless of how many sources may reach it.  The
check itself---load the filter word, hash the caller SID, and
test the corresponding bit---adds only a few instructions per
target.  Compared to Bratter~\cite{bratter}, BRL-Func incurs
moderately higher overhead than Bratter\_FS (0.46\% vs.\ 0.15\%)
due to the Bloom filter metadata storage.  However, when
CFG-level protection is enabled, this design advantage becomes
clear: Bratter may insert up to four \texttt{cbtag} instructions
at a single target to encode multiple tag values, and this
per-target cost accumulates in benchmarks with dense control
flow---Bratter\_Both reaches 1.20\% on average and 11.33\% on
\texttt{cover}.  BRL-CFG remains at 0.52\% mean, less
than half of Bratter\_Both, while providing finer-grained
source-level authorization.

Bloom filter metadata in \texttt{.rodata} contributes an additional mean of 86~bytes (BRL-Func) and a median of 264~bytes (BRL-CFG) per benchmark. As a fraction of total binary size (\texttt{.text} $+$ \texttt{.rodata}), this amounts to a mean of 0.68\% and 4.15\% respectively, though the CFG figure is dominated by \texttt{cover} (6{,}440~bytes for 181 protected targets); the median CFG \texttt{.rodata} growth is 264~bytes. We note that Bratter~\cite{bratter} stores authorization state in a dedicated CSR and incurs no binary metadata cost, making this a structural rather than implementation difference.

\begin{table}[htbp]
\centering
\caption{Code size overhead and CFI density for selected BEEBS benchmarks.
\texttt{.text OH} = static text section growth;
\texttt{CFI density} = (BRL+BLD executed) / total instructions.}
\label{tab:codesize}
\small
\setlength{\tabcolsep}{4pt}
\begin{tabular}{l rr rr}
\toprule
& \multicolumn{2}{c}{\texttt{.text} OH (\%)} & \multicolumn{2}{c}{CFI Density (\%)} \\
\cmidrule(lr){2-3} \cmidrule(lr){4-5}
Benchmark & Func & CFG & Func & CFG \\
\midrule
cover        & 0.43 & \red{7.85}  & 0.024 & \red{8.051} \\
trio-snprintf& 1.23 & 2.30        & 0.683 & \red{1.307} \\
trio-sscanf  & 0.67 & 1.88        & 0.865 & \red{1.213} \\
picojpeg     & 0.23 & 1.33        & 0.001 & 0.114 \\
lcdnum       & 0.55 & 1.26        & 0.232 & \red{2.489} \\
mergesort    & 1.20 & 0.92        & 3.008 & 2.976 \\
duff         & 0.54 & 0.83        & 0.046 & \red{1.812} \\
miniz        & 0.31 & 0.76        & 0.588 & 0.409 \\
template     & 0.58 & 0.42        & 0.931 & 0.930 \\
bs           & 0.56 & 0.41        & 0.391 & 0.394 \\
wikisort     & 0.23 & 0.25        & 3.239 & 3.226 \\
\midrule
Median (all 81)            & 0.50 & 0.37 & 0.008 & 0.008 \\
\textbf{Mean (all 81) }   & \textbf{0.46} & \textbf{0.52} & \textbf{0.147} & \textbf{0.307} \\
\bottomrule
\end{tabular}
\end{table}

\subsection{Performance Overhead}
\label{sec:eval-performance}


\begin{figure}[htbp]
\centering
\includegraphics[width=\textwidth]{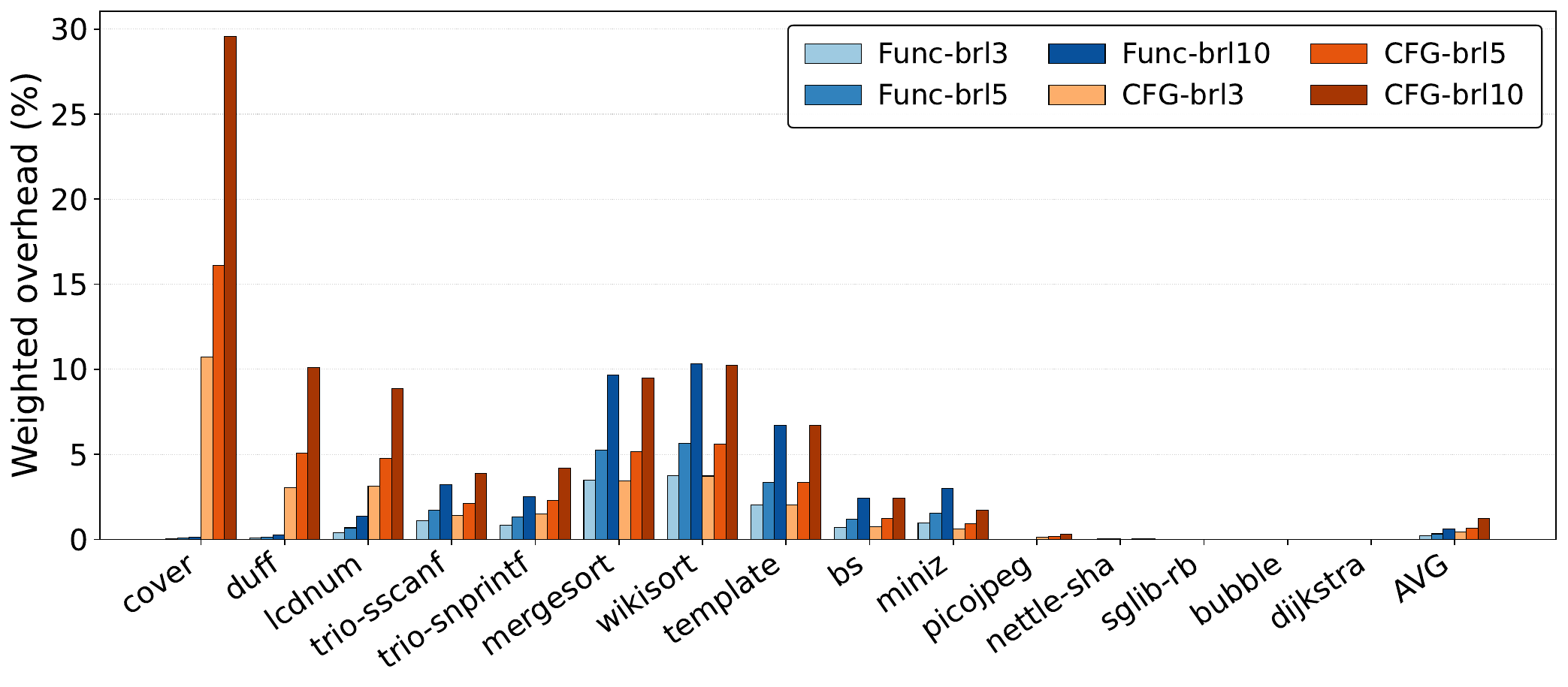}
\caption{Weighted performance overhead under BRL-Func and BRL-CFG
policies with \red{3-cycle, 5-cycle, and 10-cycle} \texttt{brl}
latency assumptions and a 1-cycle \texttt{bld}.}

\label{fig:perf}
\end{figure}

The execution time overhead is estimated using Spike's instruction
histogram multiplied by per-class cycle weights, following the
weighted-cycle methodology of Bratter~\cite{bratter}, which assigns
2~cycles to each \texttt{sbtag}/\texttt{cbtag} instruction.  We
assign 1~cycle to ALU and \texttt{bld} (immediate-to-CSR write), 2 to
taken branches, 3 to load/store, and 10 to \texttt{ecall}.
For \texttt{brl}, we evaluate three latency points---\texttt{brl3},
\texttt{brl5}, and \texttt{brl10}. We detail introduce these differences in Section~\ref{sec:discussion-latency}.
All 81~benchmarks achieve zero \textsc{Fail} across both policies
\red{under each latency model}, confirming that the instrumented
legitimate executions do not trigger enforcement failures.

Figure~\ref{fig:perf} shows the weighted overhead for selected
benchmarks.  The mean under \texttt{brl3} is \red{0.210\%}
(BRL-Func) and \red{0.421\%} (BRL-CFG), \red{rising to 0.331\% and
0.651\% under \texttt{brl5}, and to 0.633\% and 1.228\% under the
conservative \texttt{brl10} model---in all three cases well below}
Bratter\_Both's reported 5.99\% average.  Because Bratter and BRL
enforce different policy abstractions, this comparison is best
viewed as a same-benchmark reference point rather than a strict
apples-to-apples hardware comparison.  This gap arises because
Bratter's \texttt{cbtag} executes at every basic block entry
traversed at runtime, whereas our \texttt{BLD}/\texttt{BRL} pair
fires only on actual indirect transfers, which occur far less
frequently.

The one notable exception is \texttt{cover} under BRL-CFG, which
reaches \red{10.74\%} under \texttt{brl3}, \red{16.12\% under
\texttt{brl5}, and 29.57\% under \texttt{brl10}}, due to three switch
statements (120, 50, and 10 cases) now protected against JOP
attacks.  The \texttt{brl3} figure closely matches Bratter's
reported 11.1\% BR overhead for the same benchmark~\cite{bratter},
validating that our indirect jump instrumentation achieves
comparable coverage.  Outside \texttt{cover}, no benchmark exceeds
\red{3.8\% under \texttt{brl3}, 5.7\% under \texttt{brl5}, or 10.4\%
under \texttt{brl10}} for either policy.  For benchmarks without
indirect jumps (e.g., \texttt{dijkstra}, \texttt{bubblesort},
\texttt{huffbench}), BRL-Func and BRL-CFG produce identical
overhead, as the CFG-mode jump protection adds no instructions when
no indirect jumps are present.  Table~\ref{tab:perf_overhead}
provides the aggregate statistics.  \red{Because Spike is a
functional simulator, the dynamic instruction histogram is
independent of cycle-weight assumptions, so overhead under any
\texttt{brl} latency can be obtained by linear recomputation from
the same trace without re-executing the benchmarks.}

\begin{table}[htbp]
\centering
\caption{Performance overhead summary across all 81 BEEBS benchmarks under three brl latency models.}
\label{tab:perf_overhead}
\begin{tabular}{lrrrrrr}
\toprule
 & \multicolumn{2}{c}{brl3 (\%)} & \multicolumn{2}{c}{brl5 (\%)} & \multicolumn{2}{c}{brl10 (\%)} \\
\cmidrule(lr){2-3} \cmidrule(lr){4-5} \cmidrule(lr){6-7}
Statistic & Func & CFG & Func & CFG & Func & CFG \\
\midrule
Mean   &  0.210 &  0.421 &  0.331 &  0.651 &  0.633 &  1.228 \\
Median &  0.013 &  0.014 &  0.021 &  0.023 &  0.042 &  0.046 \\
Max    &  3.752 & 10.743 &  5.633 & 16.122 & 10.333 & 29.569 \\
\bottomrule
\end{tabular}
\end{table}

\subsection{Equivalence Class Size}
\label{sec:eval-ec-size}
 
The equivalence class (EC) of an indirect call site is the set of
target functions that the CFI policy permits it to invoke.  A smaller
EC means the attacker has fewer gadgets available after hijacking a
function pointer, so EC size is a widely used security metric for
CFI schemes~\cite{eccut}. For Branch Landing, a smaller EC size also means more concrete 
source-target mapping.

Table~\ref{tab:ec-size} reports the compile-time EC statistics
across all 81 BEEBS benchmarks.  Under \textsc{BRL-Func}, the 
average EC size is 1.96 targets per call site; under
\textsc{BRL-CFG}, this drops to 1.32, a \textbf{32.5\% reduction}.
The improvement is consistent: BRL-CFG tightens the EC in every
benchmark for which BRL-Func reports an EC larger than~1.
 
The reduction stems from two complementary effects.  First, the
pointer-analysis-based policy discovers more indirect call sites than
type matching alone (mean 7.47 vs.\ 5.37), which distributes the
same set of target functions over a finer-grained set of call sites.
Second, the analysis prunes spurious targets that type matching
conservatively admits, yielding smaller per-site ECs.  Notably,
\texttt{sglib-rbtree} and \texttt{sglib-dllist} achieve a 50\%
EC reduction (from 2.0 to 1.0), meaning the CFG policy narrows each
call site to a single valid target---the theoretical optimum.  Other
complex benchmarks also show significant reductions:
\texttt{sglib-hashtable} and \texttt{sglib-listinsertsort} (42.9\%),
\texttt{picojpeg} (28.6\%), and \texttt{mergesort} (18.2\%).
 
The largest EC observed across both policies is~9 (in
\texttt{wikisort} and \texttt{mergesort}), which is driven by a
sorting-routine dispatch table.  Even in this worst case the EC
remains small enough to substantially limit an attacker's choice of
targets compared to an unprotected binary.

Among related CFI schemes, ECCut~\cite{eccut} reports a 94.8\%
average EC reduction over LLVM-CFI on SPEC~CPU benchmarks.
However, ECCut operates on large x86 applications where
type-based policies produce ECs of tens to hundreds of targets,
leaving substantial room for refinement.  In contrast, embedded
benchmarks in BEEBS inherently have fewer indirect calls and
smaller ECs---our BRL-Func baseline already averages only 1.96
targets per call site.  Despite this limited headroom, BRL-CFG
still achieves a 32.5\% reduction to 1.32, demonstrating that
pointer-analysis-based refinement remains effective even in the
embedded domain.  Bratter~\cite{bratter} does not report EC
size metrics; its 8-bit tag register provides a fixed label
space that cannot express per-call-site target restrictions
without consuming additional tag bits, fundamentally limiting
its security granularity compared to our Bloom filter approach.
As discussed in Section~\ref{sec:background}, ECCut's type
refinements are complementary to BRL's membership-based enforcement
and could be integrated as a policy input to further reduce
equivalence class sizes. HPCFI~\cite{kasten2024hpcfi} reports similarly large
reductions on SPEC CPU through integrated static analyses; like
ECCut, these analysis-level refinements are orthogonal to BRL's
enforcement-level contribution and could be adopted as policy inputs
within the same framework.

\begin{table}[t]
  \centering
  \caption{Compile-time equivalence class (EC) size for selected
    benchmarks.  \textit{Call sites} = number of static indirect call
    sites; \textit{Avg EC} = mean targets per call site;
    \textit{Max EC} = largest single EC.}
  \label{tab:ec-size}
  \small
  \setlength{\tabcolsep}{3pt}          
  \resizebox{\columnwidth}{!}{
  \begin{tabular}{l rrr rrr r}
    \toprule
    & \multicolumn{3}{c}{\textsc{BRL-Func}}
    & \multicolumn{3}{c}{\textsc{BRL-CFG}}
    & EC \\
    \cmidrule(lr){2-4}\cmidrule(lr){5-7}
    Benchmark
      & Sites & Avg & Max
      & Sites & Avg & Max
      & Red.\,(\%) \\
    \midrule
    trio-snprintf   & 38 & 1.11 & 2 & 40 & 1.05 & 2 &  5.0 \\
    wikisort        & 34 & 1.35 & 9 & 36 & 1.28 & 9 &  5.6 \\
    trio-sscanf     & 17 & 1.24 & 2 & 20 & 1.05 & 2 & 15.0 \\
    miniz           & 16 & 1.75 & 3 & 19 & 1.47 & 3 & 15.8 \\
    nettle-aes      & 12 & 2.00 & 3 & 14 & 1.71 & 3 & 14.3 \\
    nettle-sha256   & 12 & 2.00 & 3 & 14 & 1.71 & 3 & 14.3 \\
    mergesort       &  9 & 2.33 & 9 & 11 & 1.91 & 9 & 18.2 \\
    picojpeg        &  5 & 1.80 & 2 &  7 & 1.29 & 2 & 28.6 \\
    sglib-hashtable &  4 & 2.00 & 2 &  7 & 1.14 & 2 & 42.9 \\
    sglib-rbtree    &  4 & 2.00 & 2 &  8 & 1.00 & 2 & 50.0 \\
    sglib-dllist    &  4 & 2.00 & 2 &  8 & 1.00 & 2 & 50.0 \\
    \midrule
    \textbf{All 81} & \textbf{5.37} & \textbf{1.96} & \textbf{9}
                    & \textbf{7.47} & \textbf{1.32} & \textbf{9}
                    & \textbf{32.5} \\
    \bottomrule
  \end{tabular}
  }
\end{table}

\subsection{Runtime Verification and Protected Targets}
\label{sec:eval-protected-targets}

BRL-Func protects only indirect-call targets (i.e., address-taken function
entries), whereas BRL-CFG additionally protects indirect-jump targets such as
switch jump-table destinations and computed-goto labels. As a result,
BRL-Func protects a mean of 2.3 targets per benchmark, while BRL-CFG increases
this number to 4.0; the increase is especially pronounced for switch-heavy
benchmarks such as \texttt{cover} (2 to 181), \texttt{picojpeg} (3 to 15), and
\texttt{trio-snprintf} (3 to 13).

We validate correctness by executing all 81 BEEBS benchmarks on the Spike ISA
simulator, which records each \texttt{brl} outcome as Pass, Skip, or Fail.
Across all benchmarks and both policies, the Fail count is zero. Under
BRL-CFG, the aggregate Pass count increases from 124.1\,M to 128.7\,M, while
Skip decreases from 1.84\,M to 0.51\,M, indicating that CFG-derived protection
converts previously unchecked transfers into actively verified ones.

\section{Discussion}
\label{sec:discussion}

This section discusses two design considerations that extend beyond
the core mechanism described in Section~\ref{sec:design}: 
out-of-order handling for \texttt{BRState} and the interpretation of the brl latency models.

\subsection{BRState Under Out-of-Order Execution}
\label{sec:disc-ooo}

Our prototype is evaluated on Spike, whose functional execution model
does not expose out-of-order (OoO) state hazards. On an OoO processor,
however, a straightforward global \texttt{BRState} design would require
care to prevent a speculative \texttt{bld} from overwriting the pending
authorization state of an earlier indirect transfer before the matching
\texttt{brl} commits.

This issue can be addressed with standard commit-order design choices.
For example, the SID carried by \texttt{bld} can be kept in the reorder
buffer and made visible to \texttt{brl} only at commit time, or
equivalently, \texttt{BRState} can be updated only when \texttt{bld}
commits. Under either design, squashed speculative \texttt{bld}
instructions never update architectural authorization state, and each
successful \texttt{brl} still consumes exactly one committed source SID.

Therefore, the single-use authorization invariant of Branch Landing is
compatible with OoO execution, although a full RTL design is beyond the
scope of this work. Our current evaluation focuses on the architectural
mechanism and reports performance using latency models rather than a
cycle-accurate OoO implementation.

\subsection{Latency Modeling and Implementation Considerations}
\label{sec:discussion-latency}

Branch Landing performs membership checking with a fixed number of
hash probes for a chosen filter configuration, so the modeled check
latency does not scale with the size of $\text{AllowedSources}(T)$,
but the absolute latency of \texttt{brl} depends on how its metadata
path is implemented.  Our evaluation uses \red{three latency
models---\texttt{brl3}, \texttt{brl5}, and \texttt{brl10}}---to
represent \red{optimistic, realistic, and conservative}
implementation points rather than a finalized RTL cost.

\red{A \texttt{brl} execution conceptually performs four operations:
(i) validate \texttt{BRState.valid}; (ii) resolve the per-target
filter descriptor $(\textit{base}, m)$ from the \texttt{.rodata}
descriptor table using the immediate $SID_T$ operand; (iii) compute
$k$ Bloom filter hash positions; and (iv) load the corresponding
filter words and AND-reduce the sampled bits.  We estimate each step
from Bloom filter hardware and cache literature.
\emph{Descriptor resolution} is a read from a small read-only table
keyed by $SID_T$.  A dedicated descriptor cache analogous to a
branch target cache can serve the lookup in 1~cycle; without it,
the descriptor falls into the L1 D-cache, incurring the
$\sim$3-cycle hit latency typical of in-order RISC-V cores such as
CVA6~\cite{zaruba2019cost}.
\emph{Hash computation} uses double hashing~\cite{kirsch2006less}
to derive $k$ filter positions from two base hashes $h_1$ and $h_2$;
a simple universal hash such as H3~\cite{carter1977universal} or the
ultra-low-latency Xoodoo-NC~\cite{sateesan2022hardware} completes both
base hashes combinationally in a single cycle.
\emph{Filter bit probes} require one memory read per position;
hardware Bloom filter studies~\cite{sateesan2022hardware,harwayne2009fpga}
report 1--2 cycle BRAM access when the filter is banked, while a
general-purpose L1 D-cache read inherits the 1--3 cycle cache-hit
latency~\cite{fuguet2023hpdcache,zaruba2019cost}.
\emph{AND reduction} and \texttt{BRState} clearing are a single
cycle of combinational logic.}

\red{Under the \texttt{brl3} model we assume a descriptor-cache
hit, a compact filter stored adjacent to its descriptor, and
parallel hash/filter access (cycle~1: descriptor lookup and valid
check; cycle~2: parallel hash computation and filter word access;
cycle~3: AND-reduce and commit).  The \texttt{brl5} model drops the
descriptor cache and serves the descriptor from the L1 D-cache,
overlapped with a 1-cycle hash and a 1-cycle filter probe plus
AND/commit.  The \texttt{brl10} model corresponds to a fully
serialized implementation with no dedicated optimizations:
3~cycles for descriptor load, 1 for hash, 3 for filter word load,
1 for AND, and 2 for \texttt{BRState} commit and pipeline refill.}

Across \red{all three} models, the qualitative conclusions of our
evaluation remain unchanged: Branch Landing continues to incur low
estimated runtime overhead while preserving source-aware
authorization.  A full cycle-accurate RTL study of these
implementation choices is left to future work.
\section{Conclusion}
\label{sec:conclusion}

This paper presents Branch Landing, a forward-edge CFI framework for RISC-V that introduces Bloom filter-based source authorization to close the source-domain gap left by existing type-based and tag-based mechanisms. Two lightweight ISA extensions, \texttt{bld} and \texttt{brl}, propagate and validate source Section Identifiers through a dedicated \texttt{BRState} register, achieving fixed-probe membership verification whose latency is independent of the authorized-set size, scaling to an arbitrary number of authorized callers. The policy-agnostic design allows both type-based and CFG-derived authorization to be instantiated within the same hardware substrate. 
Evaluation on 81 BEEBS benchmarks confirms mean runtime overheads below 0.5\% under the primary latency models, with the conservative 10-cycle sensitivity point remaining below 1.3\%, a 32.5\% equivalence class reduction under the CFG-derived policy, and zero enforcement failures on the exercised benchmark executions.

%
%
%

%
%
%
%
\bibliographystyle{splncs04}
\bibliography{ref}

\end{document}